\documentclass[11pt]{article}
\pdfoutput=1
\usepackage{moriond, epsfig}
\usepackage[latin1]{inputenc}
\usepackage[OT1]{fontenc}

\begin{document}
\vspace*{4cm}
\title{THE BRAIN EXPERIMENT \\ -A BOLOMETRIC INTERFEROMETER DEDICATED TO THE CMB B-MODE MEASUREMENT-}

\author{Romain CHARLASSIER$^{1}$ on behalf of the BRAIN Collaboration}

\address{$^1$Laboratoire APC - Univer\textbf{}sit\'{e} Paris 7 Denis Diderot - CNRS - IN2P3 \\ 10, rue Alice Domon et L\'{e}onie Duquet - 75205 Paris, France}

\maketitle\abstracts{
We present the BRAIN Experiment, a project of B-mode experiment using a novel technology, bolometric interferometry. This technique is a promising alternative to direct imaging experiments since it combines the advantages of interferometry in terms of systematic effects handling and those of bolometry in terms of sensitivity. We briefly introduce some of the bolometric interferometry key concepts and difficulties. We then give the specifications of the BRAIN future detector. A first module of the final instrument is planned to be installed at D\^ome C in 2010. We hope to constrain a tensor to scalar modes ratio of 0.01 with nine modules and one effective year of data. BRAIN is a collaboration between France, Italy and United Kingdom.}

\section{Introduction - The B-mode Quest}
%The idea of an epoch of inflation at early times during which the universe experienced accelerated expansion can solve simultaneously several questions in cosmology. Quantum fluctuations during this period lead to the creation of the density perturbations that are observed today as the large-scale structure of the universe and as Cosmic Microwave Background anisotropies. The same mechanism should also generate primordial gravitational waves of cosmological size. Their observation would give clear evidence for inflation and provide a direct measurement of the relevant energy scale (most likely associated with Grand Unification at $10^{15}-10^{16} \ GeV$, just $10^{-38}$ seconds after the Big Bang). The CMB offers a unique way to observe these primordial gravitational waves by their unmistakable imprint in the polarization anisotropies. The CMB polarization map can indeed be decomposed into a non-local base of two components, a scalar one denoted E and a pseudoscalar denoted B. On large angular scale (spherical harmonics multipoles included between 50 and 200), only tensor modes originating from primordial gravitational waves should have induced B-modes. The amplitude of the B-modes signal is however expected to be much lower than the E-mode one -the present constraint~\cite{Dunkley:2008ie} on the tensor to scalar ratio modes is $r < 0.43 \ (95\% \ CL)$; thus, their detection -sometimes presented as the Holy Grail Quest of nowadays cosmology- is at least a tremendous experimental challenge.

The amplitude of the B-mode signal is expected to be much lower than the E-mode one -the present constraint~\cite{Dunkley:2008ie} on the tensor to scalar ratio modes is $r < 0.43 \ (95\% \ CL)$; thus, its detection -sometimes presented as the Holy Grail Quest of nowadays cosmology- is at least a tremendous experimental challenge. In addition to an exquisite statistical sensitivity (huge number of feedhorns and bolometers required), future experiments will need an excellent quality of foreground removal (multiwavelength detectors required) and an unprecedented control of systematics. Systematic effects are an important field of study in themselves~\cite{Hu:2002vu}. Instrument-induced effects include for instance beam errors (mismatching and cross-polarization), gain errors (pointing and detector miscalibration), coupling (due to instrumental polarization, misalignment of polarization angles), as well as polarized sidelobes due to front optics or atmospheric polarization for ground-based experiments (this is not an exhaustive list).

\section{A bolometric interferometry primer}
\subsection{Imagers versus interferometers for the B-mode Quest}
An important remark is that most future experiments expecting to detect B-mode (EBEX~\cite{Oxley:2005dg}, QUIET~\cite{Samtleben:2008rb}, PLANCK~\cite{Pl:2006uk}, CLOVER~\cite{Taylor:2006jw}, etc.) will be imaging ones where all the same instrumental error sources will lead to the same systematic effects.  An alternative approach to the imaging systems is interferometry. Two interferometers, DASI~\cite{Leitch:2004gd} and CBI~\cite{Contaldi:2002mi}, have been the first experiments to detect E-mode polarization of the CMB. While imagers measure maps of the CMB (power spectra are then computed from them), interferometers directly measure Fourier components of the Stokes parameters. Difficult systematic problems that are inherent to total power and differential measurements are absent in a well-designed interferometer. The absence of optics in front of primary feedhorns reinforces this advantage. At least one can say that same instrumental error sources than the imaging ones lead to different systematic effects~\cite{Bunn:2006nh}. For instance, in an interferometer mismatched beams do not lead to conversion of temperature into polarization. On the other hand, any coupling between the orthogonal polarizations entering different antennas will produce a spurious correlated signal.

Unfortunately, the sensitivity of current "classical" interferometers is intrinsically limited because they are pairwise interferometers constituted by coherent receivers: the electromagnetic signals collected by two feedhorns are amplified (HEMT amplifiers in the case of DASI and CBI) and mixed to lower frequencies before being correlated by pair. Even if the amplifiers technology is in constant evolution~\cite{Samtleben:2008rb}, amplifiers are intrinsically less sensitive than cooled bolometers (at least for the 90 to 300 GHz range). Bolometers are incoherent detector that cannot be used in classical radio-interferometer designs. Above all, the hardware complexity of such interferometers grows as the square of their number of feedhorns (for $N$ feedhorns they need, $N(N-1)$ correlators) and thus, it is very difficult~\cite{Bock:2006yf} to build a coherent radio-interferometer for a number of receivers larger than a dozen (13 for DASI and CBI), knowing that typically, hundreds and even maybe thousands of receivers will be required for the B-mode detection. The goal of bolometric interferometry is therefore to combine the advantages of interferometry with the sensitivity advantages of large bolometers arrays. As we will see, it is based on additive interferometry as opposed to multiplicative interferometry used for coherent systems. 

\subsection{Bolometric Interferometry : basic concepts and expected difficulties}
The basic observables in CMB interferometry are the visibilities : the Fourier transform through the beams of the observed sky field (the power spectrum is the square modulus of these visibilities in the flat sky approximation). The basic idea of bolometric additive interferometry is that the correlation between two microwave signals can be done by the cooled bolometers themselves : the bolometers act as "square-law" devices, squaring the sum of the collected electromagnetic signals, the cross terms being the visibilities. Hence, a bolometric interferometer is composed of a 2-D array of feedhorns directly observing the sky (without front telescope). Each couple of feedhorns defines a baseline matching a multipole l. Each feedhorn is coupled to an ortho-mode transducer (OMT) in order to separate the two polarization. The collected signals are added together by a beam combiner. The beam combination can occur in a guided-wave structure such as waveguide or microstrip transmission lines (Butler combiner) or can be made optically as we will see (Quasi-Optical combiner). %We will discuss below how Butler and Quasi-Optical combiners work. 
The outputs of the combiner are then coupled to bolometers, which square the sum of all the feedhorns signals. Each bolometer signal is then a linear combination of the pursued visibilities. Hence, multiplexing is required to recover the visibility from each baseline. Controlled phase-shifters, located between the antennas and the beam combiner allow this multiplexing : phase-shift sequences are realized, and the problem of visibilities recovery can be solved. 

Practically, bolometric interferometry is not so simple. We have identified two main issues. The first is a hardware complexity one. Waveguide -or even strip line- Butler combiners are technologically very difficult to build for a large number of inputs. The combiner must at least have as many output channels as input ones, in order to conserve the input power. One has to imagine that all the input signals have to be divided and combined in such a way that every output signals is a linear combination of all the input ones. Besides, because there are no amplifiers in this system, the losses between the horn inputs and the detectors must be small. The waveguide or strip line technology of such a Butler combiner with a large number of inputs (typically $10^2$) simply does not exist yet. A solution to this issue is the Quasi-Optical Combiner conceived for the MBI~\cite{Timbie:2006} prototype : in this design, described below, the beam combination is performed optically. The second issue is a sensitivity one. We think, as it will be explained in a paper in preparation~\cite{Charlassier:2008} that the sensitivity of a bolometric interferometer critically depends on the way sequences of phase-shifts are performed in order to multiplex all the measured visibilities. The main conclusion of our study is that, with well chosen phase-shifts sequences, a bolometric interferometer can reach a sensitivity roughly equivalent to that of an imaging experiment built with the same number of horns and bolometers.

\section{The BRAIN experiment}
\subsection{The QOI design}
\begin{figure}
\centering
\includegraphics[scale=0.45]{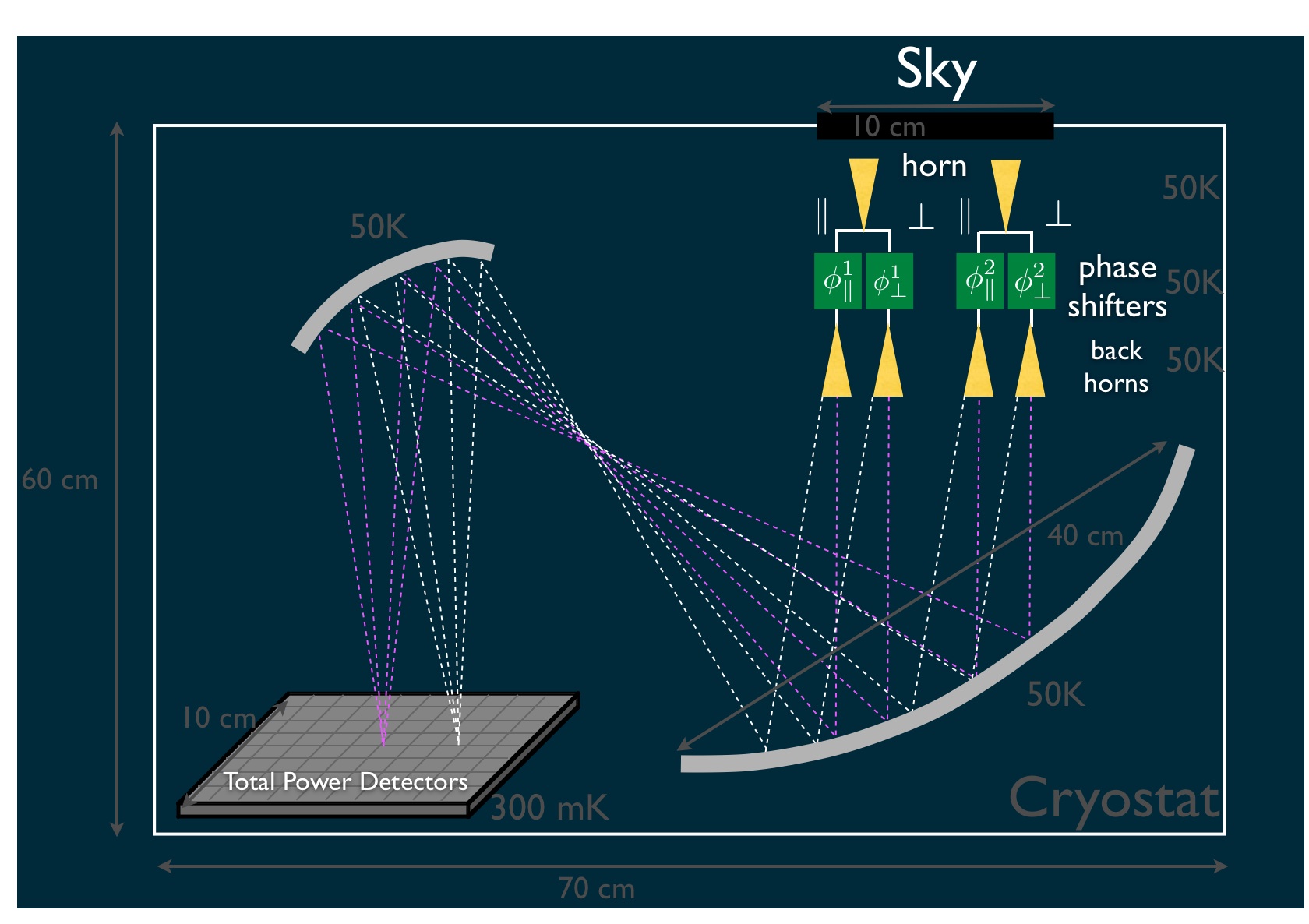} 
\caption{A possible design of the Quasi-Optical Interferometer for BRAIN}
\label{qoi}
\end{figure}
Figure \ref{qoi} shows a possible design of the BRAIN instrument based on the Quasi-Optical Interferometer concept which has been introduced by the MBI collaboration~\footnote{MBI~\cite{Timbie:2006} is the other bolometric interferometric project which already has a prototype (QOI design) with 4 horns now undergoing antenna pattern tests at the Pine Bluff Observatory in Wisconsin.}. The sky is directly observed (without front optics) by primary feedhorns. The two polarizations are separated by Ortho-Mode Transducers, and then phase-shifted by controlled phase-shifters. Secondary back horns then reemit the signals towards an off-axis telescope. All the system is installed in a cryostat cooled to 50 K. An array of bolometers, cooled to 300 mK by the second stage of the cryostat, is located in the focal plane of the telescope. Assuming back to back horns are acting as secondary sources, two signals coming from the same point on the sky but from different horns form interference patterns on the bolometer array. The Quasi-Optical beam combiner is thus equivalent to a Butler combiner : each pixel in the bolometer plane measures a linear combination of all the visibilities, whose coefficients depend on the pixel position : this is a kind of spatial multiplexing. Controlled phase-shifters also allow time-domain multiplexing. As we have already said, recovering the visibilities is not straightforward : we will not discuss it here, but the sensitivity of the detector critically depends on the multiplexing scheme~\cite{Charlassier:2008}. 

\subsection{The D\^ome C site}
Our goal is to install the BRAIN instrument in Antarctica, in the French/Italian Concordia Station located at the D\^ome C site (3233 m altitude). This site has many advantages for millimetric astronomy : a very low brightness temperature of its atmosphere (around 14 K) and an excellent transmission both due to its very low precipitable water vapor level, an exceptional atmospheric stability within the polar vortex, and a low sun set on the horizon. Furthermore, these favorable experimental conditions are available most of the year. 
%Preliminary atmospheric characterization of the site indicates that one year of observation from D\^ome C for CMB polarization measurement is roughly equivalent to two years from South Pole and five years from Atacama plateau in Chile.

\subsection{The BRAIN Program $\&$ Schedule}
The BRAIN Experiment is a collaboration between France (APC Paris, CESR Toulouse, CSNSM Orsay, IAS Orsay), Italy (Universit\`a di Roma La Sapienza, Universit\`a di Milano Bicocca) and United Kingdom (University of Manchester, University of Wales in Cardiff). Note that a joint BRAIN/MBI collaboration is under discussion. The BRAIN program is currently twofold.

First, the Pathfinder program is in charge of site testing, logistics and team preparation. We already had two campaigns (January 2006, January 2007) at the D\^ome C with the pathfinder instrument, giving us one month of observations~\cite{Polenta:2007}. Preliminary results indicate that the quality of the site is as good as expected. 

Simultaneously, we are developing the BRAIN prototype. The idea is to build a first module that will be one of the nine of the final instrument. 
This final instrument will operate at 90, 150 and 220 GHz (three modules per frequency channel). Each modules will be composed of 144 feedhorns with a beam size of $20^\circ$. The number of bolometers per module will be typically about one hundred. Recall that for an interferometer, the angular resolution is not given by the beam size of feedhorns but by the baselines size. Typically, the arrays of feedhorns will be designed in such a way that most of the baselines will observe multipoles included between 50 and 200. Note that all these specifications can evolve. The first module is planned to be installed in 2010 and the full size instrument in 2011. Our studies (paper in preparation) show that, with these characteristics, the BRAIN Experiment can constrain with a $4\sigma$ accuracy the existence of primordial gravitational waves for a tensor to scalar ratio of r=0.01 in one effective year of data. %(see figure \ref{imag_sensitivity}). 


\section*{References}
\begin{thebibliography}{99}

\bibitem{Charlassier:2008}
  R.~Charlassier, {\it et al.},
  ``An efficient phase-shifting scheme for bolometric additive interferometry''
  \textit{(in preparation)}

\bibitem{Polenta:2007}
  G.~Polenta, {\it et al.},
  ``The BRAIN experiment,''
  New\ Astron.\ Rev.\  {\bf 51} (2007)

\bibitem{Timbie:2006}
  P.~T.~Timbie, G.~S.~Tucker, {\it et al.},
  ``The EPIC and MBI,''
  New\ Astron.\ Rev.\  {\bf 50} (2006)

\bibitem{Bunn:2006nh}
  E.~F.~Bunn,
  %``Systematic errors in cosmic microwave background interferometry,''
  Phys.\ Rev.\  D {\bf 75} (2007) 083517
  [arXiv:astro-ph/0607312].

\bibitem{Hu:2002vu}
  W.~Hu, M.~M.~Hedman and M.~Zaldarriaga,
  %``Benchmark parameters for CMB polarization experiments,''
  Phys.\ Rev.\  D {\bf 67} (2003) 043004
  %%CITATION = PHRVA,D67,043004;%%

\bibitem{Leitch:2004gd}
  E.~M.~Leitch, {\it et al.} %J.~M.~Kovac, N.~W.~Halverson, J.~E.~Carlstrom, C.~Pryke and M.~W.~E.~Smith,
  %``DASI Three-Year Cosmic Microwave Background Polarization Results,''
  Astrophys.\ J.\  {\bf 624} (2005) 10
  [arXiv:astro-ph/0409357].
  %%CITATION = ASJOA,624,10;%%

\bibitem{Contaldi:2002mi}
  C.~R.~Contaldi {\it et al.},
  %``CMB observations with the Cosmic Background Imager (CBI) Interferometer,''
  arXiv:astro-ph/0210303.
  %%CITATION = ASTRO-PH/0210303;%%

\bibitem{Oxley:2005dg}
  P.~Oxley {\it et al.},
  %``The EBEX Experiment,''
  Proc.\ SPIE Int.\ Soc.\ Opt.\ Eng.\  {\bf 5543} (2004) 320
  [arXiv:astro-ph/0501111].
  %%CITATION = PSISD,5543,320;%

\bibitem{Samtleben:2008rb}
  D.~Samtleben and f.~t.~Q.~Collaboration,
  %``Measuring the Cosmic Microwave Background Radiation (CMBR) polarization with QUIET,''
  arXiv:0802.2657 [astro-ph].
  %%CITATION = ARXIV:0802.2657;%%

\bibitem{Pl:2006uk}
    [Planck Collaboration],
  %``Planck: The scientific programme,''
  arXiv:astro-ph/0604069.
  %%CITATION = ASTRO-PH/0604069;%%

\bibitem{Taylor:2006jw}
  A.~C.~Taylor  [the Clover Collaboration],
  %``Clover - A B-mode polarization experiment,''
  New Astron.\ Rev.\  {\bf 50} (2006) 993
  %%CITATION = ASTRE,50,993;%%
%
\bibitem{Bock:2006yf}
  J.~Bock {\it et al.},
``Task Force on CMB Research,''
  arXiv:astro-ph/0604101.
%  %%CITATION = ASTRO-PH/0604101;%%

\bibitem{Dunkley:2008ie}
  J.~Dunkley {\it et al.}  [WMAP Collaboration],
  %``Five-Year Wilkinson Microwave Anisotropy Probe (WMAP) Observations:
  %Likelihoods and Parameters from the WMAP data,''
  arXiv:0803.0586 [astro-ph].
  %%CITATION = ARXIV:0803.0586;%%


\end{thebibliography}
\end{document}